\documentstyle[12pt,aaspp]{article}
\begin{document}

\def\pslandinsert#1{\epsffile{#1}}
\def\eqref#1{\ref{eq#1}}
\def\e#1{\label{eq#1}}
\def\be{\begin{equation}}
\def\ee{\end{equation}}
\def\ast{\mathchar"2203} \mathcode`*="002A
\def\rslash{\backslash} \def\oforder{\sim}
\def\larrow{\leftarrow} \def\rarrow{\rightarrow}
\def\darrow{\Longleftrightarrow}
\def\defeq{\equiv} \def\lteq{\leq} \def\gteq{\geq} \def\neq{\not=}
\def\<={\leq} \def\>={\geq} \def\lsls{\ll} \def\grgr{\gg}
\def\all{\forall} \def\lub{\sqcup} \def\relv{\vert}
\def\leftv{\left|} \def\rightv{\right|}
\def\%{\char'045{}}
\def\_{\vrule height 0.8pt depth 0pt width 1em}
\def\leftbrace{\left\{} \def\rightbrace{\right\}}
\def\sectsign{\S}
\def\prop{\propto}
\newbox\grsign \setbox\grsign=\hbox{$>$} \newdimen\grdimen \grdimen=\ht\grsign
\newbox\simlessbox \newbox\simgreatbox
\setbox\simgreatbox=\hbox{\raise.5ex\hbox{$>$}\llap
     {\lower.5ex\hbox{$\sim$}}}\ht1=\grdimen\dp1=0pt
\setbox\simlessbox=\hbox{\raise.5ex\hbox{$<$}\llap
     {\lower.5ex\hbox{$\sim$}}}\ht2=\grdimen\dp2=0pt
\def\simgreat{\mathrel{\copy\simgreatbox}}
\def\simless{\mathrel{\copy\simlessbox}}
\def\spose{\rlap} \def\caret#1{\widehat #1}
\def\hat#1{\widehat #1} \def\tilde#1{\widetilde #1}
\def\limitswitch{\limits} \def\dispstyle{\displaystyle}
\def\dot#1{\vbox{\baselineskip=-1pt\lineskip=1pt
     \halign{\hfil ##\hfil\cr.\cr $#1$\cr}}}
\def\ddot#1{\vbox{\baselineskip=-1pt\lineskip=1pt
     \halign{\hfil##\hfil\cr..\cr $#1$\cr}}}
\def\dddot#1{\vbox{\baselineskip=-1pt\lineskip=1pt
     \halign{\hfil##\hfil\cr...\cr $#1$\cr}}}
\def\Abf{{\bf A}}\def\ybf{{\bf y}}\def\Ebf{{\bf E}}\def\cbf{{\bf c}}
\def\Zbf{{\bf Z}}\def\Bbf{{\bf B}}\def\Cbf{{\bf C}}\def\bfB{{\bf B}}
\def\bfq{{\bf q}}\def\bfc{{\bf c}}\def\bfy{{\bf y}}\def\bfv{{\bf v}}
\def\bfp{{\bf p}}

\def\Lya{Lyman $\alpha$}
\def\ang{\hbox{\AA}}
\def\calN{{\cal N}}
\def\Mbar{{\overline M}}

\title{PROPERTIES OF HIGH-REDSHIFT\\
       LYMAN ALPHA CLOUDS\\
       II. STATISTICAL PROPERTIES OF THE CLOUDS}

\author{William H. Press and George B. Rybicki}
\affil{Harvard-Smithsonian Center for Astrophysics,
     Cambridge, MA 02138}


\begin{abstract}
Curve of growth analysis, applied to the Lyman series absorption
ratios deduced in our previous paper, yields a measurement of the
logarithmic slope of distribution of \Lya\ clouds in column density
$N$.  The observed exponential distribution of the clouds' equivalent
widths $W$ is then shown to require a broad distribution of velocity
parameters $b$, extending up to 80 km s$^{-1}$.  We show how the
exponential itself emerges in a natural way.  An absolute
normalization for the differential distribution of cloud numbers in
$z$, $N$, and $b$ is obtained.  By detailed analysis of absorption
fluctuations along the line of sight (including correlations among
neighboring spectral frequency bins) we are able to put upper limits
on the cloud-cloud correlation function $\xi$ on several megaparsec
length scales.  We show that observed $b$ values, if thermal, are
incompatible, in several different ways, with the hypothesis of
equilibrium heating and ionization by a background UV flux.  Either a
significant component of $b$ is due to bulk motion (which we argue
against on several grounds), or else the clouds are out of
equilibrium, and hotter than is implied by their ionization state, a
situation which could be indicative of recent adiabatic collapse.
\end{abstract}


\clearpage
\section{Introduction}\label{I}

In a previous paper (Press, Rybicki, and Schneider 1993, hereafter
Paper I) we analyzed the statistics of the Lyman forest absorption, as
it is seen against the UV continuum of the background quasars
constituting the Schneider-Schmidt-Gunn high-redshift sample, whose
redshifts $z$ range from $2.5$ to $4.3$ (Schneider, Schmidt, and Gunn,
1991; hereafter SSG).  We obtained an empirical fit for mean
transmission as a function of redshift that agrees well with previous
determinations at somewhat lower redshifts, and were additionally able
to measure the mean ratios of absorption due to the lines \Lya, Lyman
$\beta$, Lyman $\gamma$, Lyman $\delta$, and possibly Lyman
$\epsilon$.  We also developed some formalism for interpreting
fluctuations in the transmission about its mean, and indicated that
these fluctuations were close to, but perhaps somewhat larger than can
be explained by a random distribution of clouds.

Paper I restricted itself to conclusions which followed directly from
the SSG data, essentially without the introduction of modeling
assumptions.  This paper broadens the analysis in several respects:

First, we will take as input to the analysis not only the SSG data as
studied in Paper I, but also some other published data, in particular
the statistical distribution of equivalent widths in the redshift
range $1.50 < z < 3.78$ reported by Murdoch et al. (1986), and the
statistical distribution of velocity parameters $b$ shown as
preliminary data by Carswell (1989; see his review for further
attribution of this published and unpublished material).

These additional data sets derive from high resolution studies that
involve the identification, counting, and detailed fitting of model
line profiles to the \Lya\ lines; we will call these results, generically,
``line counting'' measurements.  Our results in Paper I, by contrast,
did not depend on the identification of any single line, but only on
aggregate statistical properties of the Lyman forest.  In attempting
to derive uniform conclusions from a merger of these two very different
kinds of data, our challenge is to rely minimally on features of a
data set that are most susceptible to systematic error.

Second, we will eventually make some standard assumptions about the
microscopic state of the gas in the clouds.  In particular, we will
assume that the gas consists of hydrogen and helium with a primordial
cosmological abundance, and that its temperature and ionization state
are determined by local interaction (though not necessarily
equilibrium) with a posited background UV flux, conventionally, though
not necessarily, taken to be due to the continuum emission of
high-redshift quasars.

The plan of this paper is as follows: In Section 2, we apply standard
curve-of-growth analysis to the measured ratios of equivalent widths
determined in Paper I. We are able to determine (in the context of
certain parametric models) the number distribution of cloud neutral
column densities $N$ without any use of line counting measurements.
This determination is highly insensitive to the assumed distribution
of velocity parameters $b$; thus, while we cannot say anything about
$b$ from the data of Paper I alone, our characterization of the $N$
distribution is almost completely independent of any assumptions
about $b$, which is not the case in line counting measurements.

In Section 3, we show that the line counting data of Murdoch et al.
(1986) can be fit, virtually perfectly, by the column density
distribution that we determined in Section 2, if it is
independently combined with
any one of several simple model distributions for $b$.  We then
compare the measured $b$ distribution given by Carswell (1989) to the
class of acceptable model distributions.  At this point in the
analysis we are able to determine an absolute normalization on the
density of clouds, which we compare to previous published values.  We
are also able to explain the origin of the the empirical exponential
distribution of equivalent widths introduced by Sargent et al. (1980).

In Section 4 we revisit the question of fluctuations in the
absorption around its mean, last addressed by us in Paper I.  With the
detailed number distribution in both $N$ and $b$ of the previous
section, we use Monte Carlo techniques to generalize the
analytic calculations of Paper I.  We examine both the variance
and the autocorrelation (in redshift) of the absorption for signs
of an underlying two point correlation function $\xi$ in the cloud
distribution.

Sections 5 and 6 ask the question, ``Where are all the baryons?''  That is,
we investigate whether, with the distribution of $N$ and $b$ already
determined, it is possible to hide {\em all} of a plausible
cosmological density of baryons in the population of clouds that is
already known to exist at high redshifts (e.g., in the SSG sample's
Lyman forest).  We find that, with almost no model assumptions (that
is, without assuming an internal structure, or even a physical size,
for the clouds), it is possible to do so.  Indeed, the problem is
how to avoid cosmologically too many baryons in the clouds: if the
clouds are in thermal and ionization equilibrium, and if the measured
$b$ values represent thermal velocities, then neutral hydrogen fractions
are so small that the observed neutral absorption implies an overfilled
universe.

The most likely solutions are either that a significant component of
the velocity parameter $b$ is not thermal (``bulk motion
hypothesis''), or that the clouds are in an out-of-equilibrium thermal
state with respect to their ionization level, perhaps due to adiabatic
heating from recent collapse (``non-UV heating hypothesis'').  We
discuss pros and cons of the two solutions.  In either case, there is
no cosmological necessity for any nonzero Gunn-Peterson (1965) effect.

Section 7 summarizes our conclusions.

\section{Curve of Growth Analysis}\label{II}

Although curve of growth analysis is standard textbook fare, we are
applying it in an unusual range of parameters, so it is worth stating
the basic equations used.  For a line with central wavelength $\lambda_0$
and central frequency $\nu_0$,
the Doppler widths in wavelength and frequency are defined by
\def\lambdadop{\Delta\lambda_D} \def\nudop{\Delta\nu_D}
\be \lambdadop = {b\over c} \lambda_0; \qquad\qquad
    \nudop = {b \over c} \nu_0  \e{1}\ee
where $c$ is the speed of light and $b$ is the velocity parameter, given,
in the case of thermal velocities only, by
\be b = \left( {2kT\over m_H} \right)^{1/2} \e{2}\ee

For a given Lyman line of interest,
in terms of the atomic physics parameters $f$ (the oscillator strength
or $f$-value of the transition)
and $\Gamma$ (the natural decay rate of the line, $=\sum_{n\prime}
A_{nn\prime}$) one defines a Voigt parameter,
\be a \defeq {\Gamma \over 4\pi \nudop}={\Gamma \lambda_0 \over 4 \pi b}
   \e{3} \ee
and an ``integrated line optical depth,''
\be \tau_0 = {\pi e^2 \over m_e c} {1 \over \nudop} Nf =
{\pi e^2 \over m_e c}\left({N \lambda_0 f \over b}\right)  \e{4}\ee
where $N$ is the column density of neutral hydrogen
(assumed to be principally in its ground level).  In the usual
case where $a \ll 1$, the integrated line optical depth $\tau_0$
is $\sqrt{\pi}$ times the line center optical depth.

Lyman alpha clouds along the line of sight to the quasar
produce absoption lines by pure extinction, so the residual intensity
is given by
\be r(\lambda) = {F_c -F(\lambda) \over F_c} = 1-e^{-\tau_0 U(x,a)}, \e{5} \ee
where $F(\lambda)$ is the measured flux, $F_c$ is the continuum flux,
and $x \equiv (\lambda-\lambda_0)/\lambdadop$.
The normalized Voigt function is defined by
\be U(x,a) = {a\over \pi^{3/2}} \int_{-\infty}^{\infty}
   {\exp(-t^2)\over (x-t)^2 + a^2} dt \e{6}\ee
This particular normalization makes $\int_0^\infty U(x,a)\,dx=1$.

The equivalent width $W$ of the line in wavelength is then given by
\be W = \int_0^\infty r(\lambda)\,d\lambda=
   {\lambda_0 b \over c} \int_{-\infty}^\infty
     \left( 1 - e^{-\tau_0 U(x,a)} \right) dx,    \e{7}\ee
which may be computed by direct numerical quadrature.

Figure 1 shows the result of applying this prescription to the first
five lines of the Lyman series ($\alpha,\beta,\gamma,\delta,\epsilon$)
over the range $10^{13}\hbox{cm}^{-2} < N < 10^{22}\hbox{cm}^{-2}$,
and for three values of $b$ that will turn out to be of interest (30,
50, 70 km s$^{-1}$).  At the very lowest (unsaturated) and highest
(damped) column densities, one sees that the equivalent widths
have, essentially, the ratios of their respective oscillator
strengths, and they are independent of $b$.  In the intermediate
(saturated) region, the dependence on Doppler width is significant and
approximately linear in $b$.  The dependence on $N$ is linear in the
unsaturated regime, slowly varying ($\sim \log^{1/2} N$) in the
saturated regime, and varying as $N^{1/2}$ in the damped regime.
(See Spitzer 1978, \S3.4, for further details.)

Let us look now at the implications of Figure 1 to the \Lya\ clouds.
Sargent (1988) states straightforwardly an observational consensus
that (i) the typical column density of the clouds is $N\approx
10^{15}$cm$^{-2}$, and (ii) there is a power law distribution of column
densities of the form
\be f(N)dN\propto N^{-\beta}dN\e{8}\ee
with $\beta\approx 1.5$.  Indeed, there is a lively debate (into which
we are about to put an oar) on the value, and possible non-constancy,
of $\beta$ as $N$ varies over 10 orders of magnitude, from
$10^{12}-10^{22}$cm$^{-2}$ (see, e.g., Tytler 1987, Bechtold 1988,
Carswell et al. 1984, Carswell et al. 1987, Hunstead 1988, Rauch et
al. 1992, Petitjean et al. 1993).  Before proceeding,
however, we need to take note of the obvious fact that consensus
points (i) and (ii) above, if they are taken as fundamental features of
the clouds rather than as artifacts of observational selection, are
mutually incompatible: a pure power law has {\em no} typical value!

Figure 1, on the other hand, shows that the value $N\sim
10^{14}$cm$^{-2}$ {\em is} a special value observationally -- it is
where the \Lya\ line becomes saturated (for reasonable values of $b$).
Integrating the pure power law of equation (\eqref{8}) with the
equivalent width shown in Figure 1, one sees that the total absorption
will diverge for $N \gg 10^{14}$cm$^{-2}$ if $\beta <1$ and will
diverge for $N \ll 10^{14}$cm$^{-2}$ if $\beta >2$.  For $\beta$ in
the allowed range $1 < \beta < 2$, the ``typical'' cloud {\em
responsible for absorption} will have $N\sim10^{14}$cm$^{-2}$.
In this paper, and in Paper III of this series, we will several times
focus on the question of whether there is any good evidence for a
change in the properties of the {\em clouds} at $N\sim
10^{14}-10^{15}$cm$^{-2}$.  In general, we will see that the evidence
is weak.

We can thus turn with renewed interest to the question of determining
the exponent $\beta$.  In Paper I, we obtained values, and a full
covariance matrix, for the ratios
\be
{W_\beta/\lambda_\beta\over W_\alpha/\lambda_\alpha},\quad
{W_\gamma/\lambda_\gamma\over W_\alpha/\lambda_\alpha},\quad
{W_\delta/\lambda_\delta\over W_\alpha/\lambda_\alpha},\quad
{W_\epsilon/\lambda_\epsilon\over W_\alpha/\lambda_\alpha}
\e{9}\ee
We now ask whether there is any single value of $\beta$ that,
convolved with Figure 1, becomes statistically consistent with
the measured ratios.  The figure of merit is the $\chi^2$ value,
\be
\chi^2 = \sum_{i,j=1}^4 (R_i^*-R_i^0) C_{ij}^{-1} (R_j^*-R_j^0)
\e{10}\ee
where $R_i \defeq (W_i\lambda_\alpha)/(W_\alpha/\lambda_i)$,
$i=\beta,\gamma,\delta,\epsilon$, is a measured ratio from equation
(\eqref{9}), $C_{ij}$ is
the measured covariance matrix (see Paper I), and $R_i^0$ is the
theoretical value
\be
R_i^0 = { \int dN N^{-\beta} W_i(N,b)/\lambda_i \over
          \int dN N^{-\beta} W_\alpha(N,b)/\lambda_\alpha } \e{11}\ee
For a correct model, the value of $\chi^2$ should be
distributed as the standard chi-square function with 4 degrees of freedom.

Figure 2 shows the result, for three different values of $b$.  (In
Section 3 we will introduce distributions of $b$ instead of single
values.)  One sees that the exponent $\beta$ is determined to lie in
the same narrow range, approximately independent of $b$.  The values
of $\chi^2$ at the minima are quite consistent with 4 degrees of
freedom, lending credibility both to the power law model and to Paper
I's resampling estimates of the covariance matrix (which had no
knowledge of the atomic physics embodied in Figure 1).  Recalling that
a 1-$\sigma$ error estimate corresponds to a change $\Delta\chi^2=1$,
we can summarize the measured value and uncertainty (including both
statistical uncertainty and uncertainty due to the value of $b$) as
\be \beta = 1.43 \pm 0.04 \e{12} \ee
In the next section we will be more explicit about the range of $N$
that contributes to this determination, but it should be clear from
inspecting Figure 1, and from the fact that $\beta\sim 1.5$, that the
range $N\sim 10^{14}-10^{15}$cm$^{-2}$ contributes most.

It is possibly a debatable question as to whether value in equation
(\eqref{12}) is consistent with previous determinations that rely on
the counting of individual clouds.  Hunstead et
al. (1987a; see also Hunstead 1988)
obtain $\beta=1.57\pm 0.05$ fitting to a range $13.25 <
\log_{10} N < 16$, and find that a single power law is a good fit.
This contrasts with Carswell et al. (1987), who found evidence of a
break at $\log_{10}N\approx 14.35$.  Petitjohn et al. (1992) report
$\beta=1.49\pm 0.02$ as the best single power law fit over the range
$13.7 < \log_{10} N < 21.8$.  However, they obtain a somewhat steeper
value $\beta=1.83\pm.06$ for a fit restricted to $13.7 < \log_{10} N <
16$.

Considering the completely different nature of our method (not
counting clouds clouds, and using Lyman lines higher than $\alpha$),
the fact that our method is weighted differently in $N$, and the
somewhat different redshift ranges, we think that all the above quoted
errors should be taken with a grain of salt.  In favor of our method
is the fact that it is not susceptible to observational biases in the
efficiency with which one identifies, in noise, the very different
profiles of a weak line (at $N=10^{13}$cm$^{-2}$, say) and a saturated
line (at $N=10^{16}$cm$^{-2}$, say).

It must quickly be said, however, that Paper I's measured ratios of
equivalent widths, taken by themselves, do {\em not} rule out broken
power law models \`a la Carswell et al. (1987).  On the
contrary, Figure 3 shows the
results of fitting a broken power law model
\be
f(N) dN \propto \left\{ \begin{array}{ll}
   (N/N_0)^{-1.14} dN & N\<=N_0 \\
   (N/N_0)^{-1.82} dN & N>N_0
   \end{array} \right. \e{13} \ee
for the column density of the break $N_0$.  (The exponents in this
model come from a specific theoretical construction that will be
detailed in Paper III.  In this paper we will refer to this model
simply as ``Model X''.)  One finds that, for any given value of $b$,
Model X fits the data embarrassingly well, with $\chi^2\approx 1$ for
4 degrees of freedom.  (This is bound to happen about 1 time in 20
by chance.  The different values of $b$ are highly correlated and are
not independent embarrassments.)  The value of the break $N_0$ is
in the range $0.8 - 2 \times 10^{15}$cm$^{-2}$.

Figure 3 also shows, as the three ``narrow'' (approximate) parabolas,
the result of fitting to the (completely unrealistic) delta-function
model in which all clouds have a single column density $N_0$.  One
sees that, for each assumed $b$, there is a narrow window in $N_0$
that is marginally allowed.  We consider this an artifact, a kind of
``$\chi^2$ leak'', and expect that any slightly better (smaller
statistical errors) determination of the equivalent width ratios would
eliminate this -- and hopefully not eliminate the power law or
broken power law models.

To summarize: the equivalent width ratios measured in Paper I are
fairly powerful at fixing one parameter within a model (e.g., either a
single exponent, or the position of a single break), but not powerful
at distinguishing among different models.  In search of such
distinctions we move, in the next section, to consider other data.

\section{The Distribution of Equivalent Widths}\label{III}

A fundamental observational feature of the \Lya\ clouds, first
noted by Sargent et al. (1980),  is that
their equivalent widths $W$ are exponentially distributed (at least
for $W>0.3\AA$ or so) as
\be p(W)dW = \exp \left( -{W\over W^*}\right) {dW\over W^*} \e{14} \ee
Murdoch et al. (1986) give a compilation of data, including some from
Carswell et al. (1984) and Atwood, Baldwin, and Carswell (1985), all
normalized to a fiducial redshift $z=2.44$, and obtain a fitted value
$W^*=0.278\pm 0.018\AA$.  Figure 4 shows as filled circles the full
set of data given in Murdoch et al. (Pay no attention to the curves in
the figure, for now.)

One should particularly take note of the distinct departure of the
data from a pure exponential at small equivalent widths $W\simless 0.2$.
Jenkins and Ostriker (1991), among others, note that the apparent
upturn in the number of clouds with small equivalent widths can be
largely explained by the transition from unsaturated to saturated
lines (Figure 1).  We will want to see if ``largely'' can also mean
``completely''.

\subsection{{\em The Distribution in $b$}}

The quantities $N$ (column density) and $b$ (velocity parameter) are
physically more fundamental than $W$, which follows from them (as
Figure 1 indicates).  Since we have in hand, from Section 2, a
determination of the distribution of $N$ that is independent of all line
counting techniques, we can now try to invert the data in Figure 4
and obtain a distribution function for $b$.  That distribution can
then be compared to the distributions that are obtained by line profile
fitting (which, as we shall see, is a somewhat controversial subject).
This method of determining $b$ is not really independent of line
counting methods, since we must use the equivalent width data of
Murdoch et al. (or some other similar data set); but it does provide
an useful consistency check on some diverse data reduction methods.

There is controversy about $b$ on at least two issues, one of
which we regard as observational, the other theoretical.  The
observational issue is whether (as claimed by Pettini et al., 1990)
the distribution in $b$ is peaked at ``small'' values, with
a median $b$ value of 17 km/s, and essentially no values of $b$
greater than 30 km/s, or whether (as claimed by Carswell, 1989,
Rauch et al., 1992, and others) the distribution in $b$ is much
broader, with a mean value in the range 30-40 km/s and a tail
extending significantly higher in velocity.  The theoretical issue,
which is especially important if the $b$ distribution is in fact broad,
is whether the measured values of $b$ are indicative of thermal
Doppler velocities, from which conclusions can be drawn about the
microphysical state of the clouds, or whether $b$ includes
a signficant component due to bulk motion (in which case a host
of other theoretical questions are raised).

As with many inversion problems, it is better to fit a parametric
model than to try to invert ab initio.  To better understand the
sensitivities of the inversion, we will try two models.  The first
is a simple truncated Gaussian,
\be p(b) db \propto \left\{ \begin{array}{ll}
   \exp \left[ -{\displaystyle (b-b_0)^2\over \displaystyle
   2b_*^{2}}\right] db & b>0 \\
   0 & b<0
   \end{array} \right. \e{15} \ee
Here $b_0$ is the mode of the distribution (because of the truncation,
not the mean), while $b_*$ parametrizes the standard deviation.
The second model is a gamma distribution of the form
\be p(b) db \propto b^{(b_0/b_*) - 1} \exp (-b/b_*)db, \quad b>0 \e{16}\ee
where $b_0$ is the mean of the distribution, $b_*$ again parametrizes
the standard deviation.

We at this point make the assumption that $b$ and $N$ are
independently distributed (this important point will be discussed
further in Section 5, below) and we assume that $N$ is distributed as
a pure power law with $\beta$ given by equation (\eqref{12}).  Writing
the transformation of probabilities in the form
\begin{eqnarray}
  p(W) &=& \int dN \int db\, p(N,b)\, \delta [W-W(N,b)] \nonumber\\
       &=& \int dN \left[ {p(N,b) \over \partial W(N,b)/\partial b}
         \right]_{b=b(N,W)} \nonumber\\
       &=& \int db \left[ {p(N,b) \over \partial W(N,b)/\partial N}
         \right]_{N=N(b,W)}\e{17}
\end{eqnarray}
we perform the the integral (either one of the last two forms) using
the theoretical curve of growth function $W(N,b)$, and we compute
a $\chi^2$ goodness of fit measure to the data in Murdoch et al.,
using the error bars given there.  We then minimize $\chi^2$
for the two parameters $b_0$ and
$b_*$ in each of the two models (\eqref{15}) and (\eqref{16}).
The results, along with some values that characterize different
moments of the fitted distributions (useful later) are given in
Table 1.  The long- and short-dashed lines in Figure 4 show the
result of computing equation (\eqref{17}) for the best fitting models.
The agreement is seen to be excellent for both models.  (The $\chi^2$
per degree of freedom is significantly less than unity, suggesting
that at least some of the error bars shown in Murdoch et al. 1986 are
overestimated.)

\begin{table}
\begin{center}
\begin{tabular}{|lccccc|}  \hline\hline
Distribution & best $b_0$ & best $b_*$ & $\left< b \right>$ &
  $\sigma_b$ &
  $\left< b^{8.54} \right>^{1/8.54}$ \\
  & \multicolumn{5}{c|}{(all values in km s$^{-1}$)} \\ \hline
Gaussian & 32 & 23 & 36 & 20 & 61 \\
Gamma    & 38 & 14 & 39 & 23 & 77 \\
Carswell data & & & 36 & 18 & 70 \\  \hline
Mean and Spread & & & $37 \pm 2$ & $20 \pm 3$ & $69\pm 8$ \\ \hline\hline
\end{tabular}
\end{center}
\caption{Fitted parameters for the distribution of velocity parameters
$b$ (see text).}
\end{table}

Figure 5 shows (again as long and short dashes) the shapes of the
best-fitting model distributions.  Also shown in that figure, as a
histogram, are the data for the distribution of $b$ given in Carswell
(1989).  While Carswell cautions that the data (derived from several
sources) is preliminary, and rightly warns of the difficulty of
deriving such data from line profile fitting, we see that the basic
shape and parameter values of the Carswell data are in good agreement
with our fitted models that derive from the equivalent width distribution
of Murdoch et al. (1986).  We therefore include the Carswell data as
an additional line in Table 1.  The last line of that Table gives
mean values over the previous lines (two fitted models plus one actual
data set) and the spread of the models, which can be taken as
an indication of the model uncertainty.  (The standard errors of the
fitted estimates for $b_0$ and $b_*$ are also comparable, around
2 km/s.)

We have also integrated Carswell's data on $b$, via equation (\eqref{17})
and the $\beta$ value of equation (\eqref{12}), to get its predicted
distribution of $W$.  This is shown as the solid line in Figure 4.
Note that there are {\em no} new adjustable parameters in this calculation,
except for the overall normalization (moving the curve up or down).
One sees that the result is in excellent agreement with the Murdoch data,
virtually as good as the other two models.

As is obvious by Figure 1, it is the large $b$ tail of the
distribution that produces the large $W$ exponential tail.
While it is conceivable that the Carswell and Murdoch data sets, both
obtained by line counting techniques, share a common tendency to
misidentify blends, or lines from metal-line clouds, as
\Lya\ lines of large $W$, it seems quite
unlikely that the general shape of the
exponential tail in $W$, independently verified measured by many
observers from Sargent et al. (1980) to the present and in several
different redshift ranges, could be wholly an artifact of such
misidentifications.  We conclude that the high-$b$ tails of the
distributions in Figure 5, with $b$'s in the range 40 to 80 km/s, are
real and observationally mandated by the exponential distribution
of $W$'s.

Thus, as regards the observational controversy mentioned above,
we must be firmly on the side of a broad $b$ distribution.  In this,
our conclusion agrees with Rauch et al. (1992),
who discuss possible sources of discrepancy in the Pettini et al.
data.

Even if we exclude from consideration the long exponential tail in the
$W$ distribution, our data do not support entirely small $b$ values: In
Figure 4 we have plotted, as a dotted line, the best-fitting curve
that derives from a single value (delta function distribution) of $b$.
The best-fitting value is 33 km/s (cf. Table 1).  One sees that the
equivalent width distribution for $W<0.4$\AA\ is reproduced tolerably
well (though not as well as any of the broader distributions), but
that there is a severe shortage of large $W$ values.

In fact, the larger tail values for $W$ are produced, we find, by clouds that
have {\em both} larger $b$ values and larger $N$ values.  Figure 6
shows results from the same integrations that produced the curves in
Figure 4.  At each equivalent width $W$, we show the mean $N$ of the
clouds that contributed {\em to that value of $W$}, independent of their
values $b$.  One sees that values $N$ up to $\sim 10^{16}$cm$^{-2}$ are
directly probed by the data points in Figure 4.  Thus, Figure 4 yields
strong consistency check on the value, and constancy of $\beta$ in the
power law distribution of $N$: Its value, derived from
equivalent width ratios of the Lyman series, came primarily from
the saturation bend in Figure 1, at $N\sim 10^{14}$cm$^{-2}$. Now,
we see (Figures 4 and 6) that the correct number of clouds at
$N\sim 10^{16}$cm$^{-2}$ is obtained well within a factor of 2.

To emphasize this point, we have plotted, as the final curve in Figure
4, the result of integrating the broken power law Model X (equation
\eqref{13}) with the observed Carswell distribution of $b$.  One
sees that it falls significantly below the data {\em both} at small
equivalent widths {\em and} at large equivalent widths, demonstrating
the apparent relative constancy of $\beta$, at least
over three orders of magnitude in $N$.

Now to answer another question raised at the beginning of this section, we
can see that the upward bend in the number of clouds at small
equivalent width (Figure 4) can be explained {\em completely} as the effect of
folding the standard curve of growth (Figure 1) into a constant
$\beta$ (pure power law) model.  (This, incidentally, suggests caution in
approximating that part of the distribution as another exponential, as
in Jenkins and Ostriker 1991, since it is actually a power law that
diverges at zero.)  To summarize, we find no feature in any of the data
sets examined that conflicts with a pure power law distribution for
$N$; no data set requires, or even unambiguously indicates, a
``typical'' $N$ value of $10^{15}$cm$^{-2}$, or any other value, for
the underlying cloud population.

Nothing yet in our discussion takes a position on what we have above
labeled the theoretical controversy surrounding $b$, i.e.,
indicates whether the observed $b$'s derive purely from thermal
Doppler broadening, as opposed to an additional component due to bulk
motions.  We will return to this point below.

\subsection{{\em The Distribution of Clouds in $N,b,z$}}

\def\NO{{{\cal N}_0}}

We can make explicit at this point a ``standard'' working model for the
distribution of the \Lya\ clouds in $N$, $b$, and redshift $z$ at
high redshift: Let $N_{14}$ denote the value $N/10^{14}$cm$^{-2}$, and let
$n(N,b,z)\,dN_{14}\,db\,dz$ denote the mean number of clouds of in an
interval $dN_{14}$, $db$, $dz$. Then,
\be n(N,b,z) =  \NO (1+z)^{\gamma} N_{14}^{-\beta} p(b) \e{18} \ee
where $p(b)$, now assumed to be normalized to unity,
\be \int p(b) db = 1 \e{19}\ee
has the shape of any of the three distributions in Figure 5 (for
example, equations \eqref{15} or \eqref{16} with parameters from
Table 1).
Our best estimate of $\beta$ (equation \eqref{12}) is 1.43.  Our best
estimate of $\gamma$ (Paper I) was 2.46.  We can estimate $\NO$
by equating the observed mean transmission (Paper I) to the integral
of equation (\eqref{18}) weighted by the appropriate equivalent widths,
namely
\be 0.0037 (1+z)^{1+\gamma} = \int {(1+z)W(N,b)\over\lambda_0}  \NO
     (1+z)^{\gamma} N_{14}^{-\beta} p(b) \,dN\,db \e{20}\ee
Doing the integral numerically for the three $b$ distributions in
Figure 5, we obtain the absolute normalizations
\be  \NO = (2.57,\,2.54,\,2.63) \e{21}\ee
where the three values are using the observed (Carswell), gamma, and
Gaussian distributions, respectively.  The values are not, of course,
as accurate as the number of significant figures shown, but are given
to illustrate how little uncertainty is due to our remaining ignorance
of the $b$ distribution.  The units of equation (\eqref{21}) are
``clouds per unit redshift per unit $\Delta N=10^{14}$cm$^{-2}$
at $N=10^{14}$cm$^{-2}$''.  Most of the error in determining $\NO$
comes from the highly correlated errors in $\gamma$ and the value
$0.0037$, as described in Paper I.  For redshifts in the range $3 < z
< 4$, we estimate the overall error of equation (\eqref{21}) to be
on the order of $\pm 10$\%.

To compare with previous results by other investigators, we have also
computed numerically the number of clouds with equivalent widths
$W>0.32\AA$ (that value being conventional in the literature) per unit
redshift, a quantity conventionally denoted $N_0$.  We obtain
\be N_0 = (4.2 \pm 0.5) (1+z)^{\gamma} \e{22}\ee
(for redshifts $3 < z < 4$) which can be compared to the estimate of
Murdoch et al. (1986) of $4.06$ (no error given), and Jenkins and
Ostriker's (1991) estimate of $9\pm 3$, obtained from Murdoch's data
for $n(W)$ (the data points in our Figure 4) and their own
measurements of the \Lya\ broadband decrement $D_A$.

\subsection{{\em Upper Limits to Size Derived from Spacing}}

Knowing an absolute normalization on the distribution in $N,b,z$,
we immediately get some upper limits to the size of the clouds:
Since the number of clouds decreases rapidly with increasing $N$,
the average size of clouds with column density $\simgreat N$ must
be less than the average spacing along the line of sight of such clouds.
There is, of course, no reason to think that this bound lies close
to the true cloud sizes.

Distance along the line of sight is variously parametrized by
the redshift separation $dz$, the separation in observed wavelength
$d\lambda_{obs}$, the separation in comoving distance $dr$, or the
separation in physical (proper) distance $dr_{p}$.  The relation
among these quantities is
\be (1+z)dr_p = dr = {3000 \hbox{Mpc}\; h^{-1} \over (1+z)(1+\Omega z)^{1/2}}
    \, {\Delta \lambda_{obs}\over\lambda_0} =
     {3000 \hbox{Mpc}\; h^{-1} \over (1+z)(1+\Omega z)^{1/2}}
    \, dz \e{22a} \ee
where $h$ parametrizes the  Hubble constant $H_0$ by
$h\defeq H_0/100$ km s$^{-1}$Mpc$^{-1}$.
Integrating equation (\eqref{18}) in both $b$ and $N$, the number
of clouds with column densities greater than $N$ in a redshift interval
$dz$ is
\be dn_{>N} = {\NO \over (\beta-1) N_{14}^{\beta-1}} (1+z)^\gamma dz
    \e{22b} \ee
Equations (\eqref{22a}) and (\eqref{22b}) imply a limit on the physical
size $r_{phys}$ of clouds
\be  r_{phys} < {dr_p\over dn_{<N}} = {3000\,h^{-1}\hbox{ Mpc }(\beta-1)
    N_{14}^{\beta-1}
    \over \NO}\, {1\over (1+z)^{\gamma+2} (1+\Omega z)^{1/2} } \e{22c} \ee
Evidently the tightest limit is obtained by applying the argument at
the highest redshift and to the smallest-$N$ clouds.  In Paper I, we
saw that the value $\gamma \approx 2.46$ was justified up to a redshift
of at least $z \approx 4.2$, while Figure 6, above, shows that the
observed equivalent width distribution of clouds justifies the value
$\beta \approx 1.43$ down to at least $N = 10^{13}$cm$^{-2}$, i.e.,
$N_{14}=0.1$.  Taking these values, and $\NO$ from equation (\eqref{21})
above, we get, for $N\sim 10^{13}$cm$^{-2}$ clouds at $z\sim 4.2$,
\be r_{phys} < \left\{ \begin{array}{ll}
          120\,h^{-1}\hbox{ kpc} & \mbox{ if $\Omega=0$}\\
          50\,h^{-1}\hbox{ kpc} & \mbox{ if $\Omega=1$} \end{array} \right.
\e{22d} \ee

Sargent (1988) summarizes the more direct evidence for the size of the
clouds that comes from studies of close (or lensed) lines of sight at
somewhat lower redshifts.  It is believed that ``typical'' clouds
(which we would infer to be at $N\sim 10^{14-15}$cm$^{-2}$) have sizes
in the range $\sim$8 to $\sim$20 kpc; however, these values are fairly
uncertain.  We return to the issue of cloud size in \S5.

\subsection{{\em Where Does the Exponential Come From?}}

  We have already seen in the three best-fitting curves of Figure 4 that
the ``standard'' model of equation (\eqref{18}), with the parameter
values given, is able to reproduce the long-observed exponential tail
in the distribution of $W$, equation (\eqref{14}).  One might well
wonder how this manages to be true, since (if we momentarily exclude
the gamma distribution model for $b$) there are apparently no linear
exponentials anywhere in the model: the distribution in $N$ is power
law, the distribution in $b$ is (or can successfully be taken to be)
Gaussian, and the curve of growth varies in the saturated regime as
$\ln^{1/2} N$.  The most true, but least illuminating, ``explanation''
is simply that the numerics of integrating together these three
functional forms happens, readily, to produce something very close to
-- though not exactly -- a linear exponential in $W$.

A more illuminating, though necessarily crude, explanation can be
found in a toy analytical calculation which actually gives about the
right value for the exponential scale $W^*$ in equation (\eqref{14}):
{}From equation (\eqref{6}), we note that for small values of $\tau_0$ the
equivalent width is approximately $W =  \lambdadop \tau_0$; this is
the ``linear'' part of the curve of growth, where $W$
increases linearly with $N$.  This linear growth saturates when
the line optical depth becomes of order unity, or when
$\tau_0 = 1$.  This implies a saturation value
\be  W_{sat} =  \lambdadop \e{23}\ee
which occurs at the saturation column depth of
\be  N_{sat}={m_e c \over \pi e^2} {b \over \lambda_0 f} \e{24}\ee
For $N$ larger than $N_{sat}$, we shall approximate the ``flat'' part of
the curve of growth literally by taking the equivalent width to be
the constant $W=W_{sat}$.  We shall ignore the so-called ``square root''
part of the curve of growth, which occurs for much larger values of $N$,
as it appears from our numerical work that this part does not
influence the results for the  $W$'s we are concerned with.
Our approximate relation for the equivalent width is then
\be  W(N,b)=  \lambdadop \min(N/N_{sat},1)= A \min(N,Bb) \e{25}\ee
where the quantities $A$ and $B$ are defined by
\be  A = {\pi e^2 \over m_e c} \lambda_0^2 f; \qquad\qquad
     B = {m_e c \over \pi e^2 \lambda_0 f}. \e{26}\ee

Now starting with the first form of equation (\eqref{17}), we have
\begin{eqnarray}
  p(W) &\prop& \int dN \int db \delta [W-A\min(N,Bb)] N^{-\beta}
p(b)\nonumber\\
     &=& \int_0^\infty db\, \delta(W-ABb) p(b) \int_{Bb}^\infty dN\, N^{-\beta}
+\int_0^\infty dN\, \delta(W-AN)N^{-\beta}\int_{N/B}^\infty db\,p(b)
\nonumber\\
       &=& {1 \over AB} p\left({W\over AB}\right) \int_{W/A}^\infty dN\,
 N^{-\beta} + {1\over A} \left({W\over A}\right)^{-\beta}\int_{W/AB}^\infty
db\,
  p(b)  \e{27}
\end{eqnarray}
We next choose the Gamma distribution (\eqref{16}) to represent
$p(b)$, along with the numerical values for the constants $b_0$ and
$b_{*}$ given in Table 1.  Then
\be  p(W)\prop {1\over (\beta-1)AB^{\alpha}}
      \left({W\over A}\right)^{\alpha-\beta}
 e^{-W/W_{*}} + {b_{*}^{\alpha} \over A} \left({W \over A}\right)^{-\beta}
              \Gamma(\alpha, W/W_{*})  \e{28}\ee
where $\alpha \equiv b_0/b_{*} = 2.7$ and $\Gamma$ denotes the incomplete
gamma function.  We have also defined the quantity
\be
   W_{*} \equiv ABb_{*} = {\lambda_0 \over c} b_{*} \e{28a}\ee
For Lyman alpha (assuming a mean redshift factor of $(1+z) \approx 4.5$)
this implies $W_{*} = 0.18 b_6$ \AA, where
$b_6$ is the value of $b_{*}$ in units of $10^6$ cm s$^{-1}=10$ km s$^{-1}$.

In the asymptotic regime $W/W_{*} \gg 1$, the incomplete gamma function
can be replaced by its asymptotic value, which yields (omitting
some overall factors),
\begin{eqnarray}
  p(W) &\prop&   \left({W/ W_{*}}\right)^{\alpha-\beta}
         e^{-W/W_{*}}  \left[ 1 + {\beta-1 \over W/W_{*}} \right] \nonumber\\
       &\sim& \left(W/W_{*}\right)^{\alpha-\beta}
         e^{-W/W_{*}}, \e{28b}
\end{eqnarray}
the first term clearly dominating.
In the opposite limit $W/W_{*} \ll 1$, the second term dominates
in equation (\eqref{28}), since $\Gamma(\alpha,W/W_{*}) \approx \Gamma(\alpha)$
and the factor $(W/A)^{-\beta}$ becomes very large.  Thus
\be  p(W) \prop (W/W_{*})^{-\beta} \e{28c}\ee

It can be seen that this toy model shows more or less the right behavior.
The large $W$ limit (equation [\eqref{28b}]),
dominated by the flat part of the curve of growth, is roughly exponential
[with some deviation induced by the factor
$(W/W_{*})^{\alpha-\beta} \approx (W/W_{*})^{1.3}$].
In the small $W$ limit, dominated by the linear part of the curve of growth,
the power law in $N$ makes its
presence felt, and the distribution has the singular behavior
$(W/W_{*})^{-\beta}$ as $W$ goes to zero.  These two behaviors are
clearly seen in figure 5.  Furthermore, the predicted value of
$W_{*}$ for the preferred value of $b_{*}=14$ km s$^{-1}$ is
$W_{*}=0.25$\AA, which, given the crude approximations made, is not too
far from the value given after equation (\eqref{14}).
It should be emphasized that the
approximations made in this calculation are {\em not} well justified,
and that the answer is justified by the accurate numerical integration,
not by this calculation.  However, this toy example does show how it
is possible for an exponential like equation (\eqref{14}), long
observed and not previously explained, to emerge
from a featureless distribution like equation (\eqref{18}).
It also explains the upturn at small equivalent widths as
a consequence of the power law distribution of column densities.

\section{Fluctuations in the Transmission}\label{IV}

In Paper I we measured the variance of the transmission $Q$ for the
Lyman forest of the SSG quasar sample, and obtained
\be
\hbox{Var}(Q)/\overline{Q}^2 = 0.06 \pm 0.01
\e{29}\ee
This value has no fundamental significance by itself -- it depends in
part on the particulars of SSG's spectroscopy, for example their slit
resolution.  In Paper I we developed an analytic model based on a
power law distribution of $N$, like equation (\eqref{18}) in this paper,
but with a fixed line width (not equivalent width) $W_0$.  The result was
\be
{\hbox{Var}(Q)\over \overline{Q}^2} = {2W_0\over\Delta} \left[ {1\over \kappa}
   (e^\kappa-1)-1\right]
\e{30}\ee
where
\be
\kappa \defeq (2-2^{\beta-1}) \ln{1\over\overline{Q}}
       \defeq (2-2^{\beta-1}) \overline{\tau}
\e{31}\ee
and $\beta$ has its same meaning as in equation (\eqref{18}).
The instrumental resolution $\Delta$ is defined in general by
\be
\Delta = \left.  \left(\int w d\lambda \right)^2  \right/ \int w^2 d\lambda
\e{32}\ee
where $w(\lambda)$ is the instrumental response to a monochromatic
line as a function of wavelength (in the same frame as $W_0$ is
defined, either observed or at the absorption redshift).
We noted in Paper I that, for the SSG data, equation (\eqref{29})
and (\eqref{30}) were in reasonably good agreement, though with perhaps
some possibility that the required $W_0$ was unreasonably large, i.e.,
that the variance might be larger than could be explained by
Poisson statistics alone.

We are now in a position to be more quantitative about whether there is
a discrepancy, and also to look not only at the variance but also at
quantities like the covariance
$\left<Q(\lambda)Q(\lambda+\delta\lambda)\right>$
which contain information on the two-point spatial correlation
function of the \Lya\ clouds.  Suppose that $\xi(s)$ is the two point
correlation function of the clouds at an observed wavelength separation $s$.
(Below, we will convert from wavelength separations to comoving
cosmological distances.)  That is, given one cloud, the probability of
finding a second cloud separated from the first by an observed absorption
wavelength difference $s$ is the mean probability times $1+\xi(s)$.

Because of finite spectral resolution of the SSG data, and the fact that
individual clouds are not counted, the quantity $\xi(s)$ is not directly
accessible to measurement.  Instead, we can measure the covariance
of the binned values of transmission,
\def\taubar{\overline{\tau}}
\def\Qbar{\overline Q}
\be \Xi_J \defeq {1\over \ \taubar^2}\,
{\left< Q_I Q_{I+J} \right> - \left< Q_I^2 \right>\over \left< Q_I^2 \right> }
\e{33}\ee
where the angle brackets average over $I$, the index of the
10\AA-separated bins of the SSG spectroscopic data.
The factor of $\taubar^{-2}$ (where $\taubar$ is the mean optical depth)
scales $\Xi$ to measure
fluctuations in optical depth (that
is, to fluctuations in number of clouds) rather than fluctuations
in transmission:  The two scalings are related by
\be Q \sim \exp(-\tau) \quad \hbox {implying} \quad {\delta\tau\over\tau}
  \sim {1\over\tau}\, {\delta Q\over Q} \e{34} \ee

If $w(s)$ is the instrumental response (as in equation \eqref{32}) above,
then $\Xi_J$ (observable) is related to $\xi(s)$ (underlying)
by
\be \Xi_J = { \int w*w(s)\, \xi(s+J\Delta\lambda) ds \over
             \left[ \int w(s)^2 ds \right] } \e{35} \ee
where $\Delta\lambda$ is the bin separation (here 10\AA), and
$w*w(s)$ is (a single function of $s$) the convolution of $w$ with
itself,
\be w*w(s) \defeq \int w(s+y) w(y) dy \e{36} \ee

There are two interesting limits for equation (\eqref{35}):
If $\xi(s)$ is slowly varying on a scale $x \gg \Delta$ (the
instrumental resolution, equation \eqref{32}), then $\Xi$ measures
$\xi$ directly at wavelength separations $J\Delta\lambda$,
\be \Xi_J \approx \xi(J\Delta\lambda)\, {\int w*w(s) ds \over
   \left[ \int w(s)^2 ds \right] } = \xi(J\Delta\lambda) \e{37} \ee
If, on the other hand, $\xi(s)$ is nonnegligable only at values of
$s$ that are $\ll \Delta$ (unresolved structure), then
\be \Xi_J \approx {\int \xi(s) ds\over \Delta}
    { \int w*w(J\Delta\lambda) ds \over \int w^2(s) ds} \e{38} \ee
so that, in particular, the unlagged ordinary variance is given by
\be \Xi_0 \approx {\int \xi(s) ds\over \Delta} \e{39} \ee
while higher values $\Xi_J$ decay as the normalized convolution of the
instrumental response with itself.  (For a square slit, this
convolution is a triangular response starting at unity and going to
zero when $s$ equals the full width of the slit.)

It is sometimes useful to think of $\xi(s)$ as having two additive
parts, a Poisson piece $\xi_P$ that is some multiple of a Dirac delta
function $\delta(s)$ at zero lag, and which describes fluctuations due
to the discreteness of the clouds, and a smooth piece $\xi_S$ that is
generated by spatial variations in the underlying mean density of
clouds.  (See Peebles, 1980, \S33--\S36, especially equation 36.7.)
Then equation (\eqref{30}), divided by $\taubar^2$, can be viewed as
calculating the strength of the delta function piece $\xi_P$ in
equation (\eqref{39}).

With this formalism in hand, we now need to upgrade equation
(\eqref{30}) to a more realistic model of cloud line statistics, in
particular to the model equation (\eqref{18}) above.  We have done
this by Monte Carlo experiment.  Repeatedly realizing equation
(\eqref{18}) with Poisson statistics along a simulated line of sight,
we have accumulated statistics on how the resulting variance varies
with mean optical depth (or with $\kappa$ in equation
\eqref{30}).  We find that the functional form of equation (\eqref{30})
is well reproduced, but with an ``effective'' value for $W_0$ (in the
observed frame, e.g., with $\Delta=25$\AA\ for the SSG data)
\be W_0 \approx (5.5\pm 1) {\left< b \right> \over c} \left<1+z\right>
     \lambda_0\e{40}\ee
The scaling with $\left< b \right>$, $\lambda_0$, and $c$
are required by dimensional analysis.  The factor $(1+z)$ puts
$W_0$ into the observed frame, for comparison with $\Delta$.  The Monte
Carlo calculations verify (at about the 20\% level) the shape of the
dependence on $\taubar$, and compute the overall constant.

Using the numerical values $\left< b\right> = 37$km/s (Table 1),
$\lambda_0 = 1216$\AA, $\Delta = 25$\AA\ (SSG), $ \left<1+z\right>=4.5$
and $\taubar = 0.67$
(Paper I, and SSG's mean redshift), equations (\eqref{30}),
(\eqref{31}), and (\eqref{40}) give
\be
\hbox{Var}(Q)/\overline{Q}^2 = 0.076
\e{41}\ee
which is in fairly good agreement with Paper I's measured value (or
equation \eqref{29} above), and which will prove to be in excellent
agreement with a slightly different way of reducing the data, below.

It appears, then, that there is no significant clumping of the \Lya\
clouds at high redshifts $z\approx 3.5$ in the zero-lag variance:
essentially all of the variance, i.e., the fluctuations in the
absorption, can be explained as Poisson fluctuations
in a randomly distributed population of clouds drawn from the
distribution of equation (\eqref{18}).  In the language introduced
above, $\xi_P$ is as expected from equation (\eqref{18}), and $\xi_S$
is consistent with zero.

What about the covariance at nonzero lags, equation (\eqref{33})?  The
solid circles in Figure 7 show the result of calculating this quantity
(or, actually, $\taubar^2$ times this quantity) for the SSG data.
Several remarks and caveats are necessary in interpreting this Figure:

It is actually quite difficult to determine the zero point of
the covariance with high accuracy.  The structure function of the
absorption data is found to increase slowly with increasing lag, as
would be expected from small errors in fitting the the continuum
slopes of the individual SSG quasars.  This causes the raw correlation
function (essentially a constant minus the structure function) to have
an ill-defined additive constant (see e.g. \S 3 of Press, Rybicki, and
Hewitt 1992).  We fix the zero point in Figure 7 by the {\em
assumption} that there should be no significant mean correlation $\xi$
on scales of 10 to 20 bins (corresponding, we will see below, to
several tens of comoving Megaparsecs): In this range, we fit for an
r.m.s. continuum slope error, and extrapolate the resulting fitted model
back to smaller lags.

The astute reader might notice that
the value at zero lag,
\be
\hbox{Var}(Q)/\overline{Q}^2 = 0.073 \pm 0.008
\e{42}\ee
is different from Paper I's value, quoted in equation (\eqref{29})
above.  The reason is the different method, just described, used to
determine the zero point.  (See Paper I for a description of the
moving average method there used to remove systematic errors due to
incorrect extrapolations of the continuum spectra of the SSG quasars.)
One would be correct in concluding that there may well be on the order
of $0.01$ in unremoved systematic errors in the data shown in Figure
7.  The error bars shown are determined by resampling of the SSG
quasars, and should accurately reflect the statistical errors, but not
the systematic errors, of the points.

The solid line in Figure 7 is the result of predicting
$\hbox{Var}(Q)/\overline{Q}^2$ from equation (\eqref{35}), a square
slit response function of width 25\AA\ (SSG's value), and assuming
Poisson distributed clouds (equations \eqref{30} and \eqref{40})
with no additional correlation function, i.e., $\xi_S=0$.
There are {\em no} new free parameters in this prediction, i.e.,
it is {\em not} normalized to the data in any way.  We see that not
only is the agreement excellent at zero lag, but also the expected
slit response is accurately reproduced at lags 1 and 2.  In bins
3 through 7 or 8, one sees a very interesting ``shoulder'' decaying
from about 0.01 (or $\xi=0.01\taubar^{-2}\approx 0.02$), to about
half this value.  While there may be some diffraction widening of the
slit response present in bin 3, it is quite unlikely that it should
extend to bin 7 (70\AA\ out).  Further, the shoulder is highly significant
in terms of the statistical errors measured by resampling.
Unfortunately, the shoulder is at about the amplitude where we
do not have confidence in our removal of systematics.  We consider
it suggestive as a marginal, but not believable, detection of a correlation
function at high redshifts and large scales.

With greater confidence we can assert a value of
twice the shoulder value as a good upper limit to the
value of the correlation function $\xi$ on scales from 3 to 7 bins.
That is, at redshift $z\approx 3.5$,
\begin{eqnarray} \xi &<& 0.04 \quad\hbox{for}\quad \Delta\lambda_{obs}
   = 30\hbox{ \AA} \nonumber \\
   \xi &<& 0.02 \quad\hbox{for}\quad \Delta\lambda_{obs}
   = 70\hbox{ \AA} \e{43}
\end{eqnarray}
The relation between $\Delta\lambda_{obs}$ and comoving distance scale
$\Delta r$ was already given in equation (\eqref{22a}).  It follows
that the 30\AA\ and 70\AA\ scales in equation (\eqref{43}) correspond
to 7.8 and 18 $h^{-1}$ Mpc, respectively, if $\Omega \approx 1$; or to 16 and
38 $h^{-1}$ Mpc, respectively, if $\Omega \approx 0$.

It is tempting to identify the apparent shoulder as the progenitor, in
the gravitational instability picture, of correlation structure that
is seen today.  In an $\Omega=1$ universe, the growth of perturbations
in the linear regime would increase $\xi$ to the present by a factor
$(1+z)^2\approx 20$, so that the scaled values of equation
(\eqref{43}) are within a plausible factor of today's observed values
(e.g., the IRAS galaxies in Saunders et al. 1992) on the corresponding
spatial scales.  Indeed, one might hope to determine the value of
$\Omega$ from such an analysis, since the growth factor of
perturbations to the present is suppressed by a known factor in lower
$\Omega$ cosmologies.

The problems with carrying out this program
include (i) the uncertain systematic errors in Figure 7; (ii) the fact
that, if simplistic comparison with today's correlation function is
justified at all, then one predicts much larger covariances at lags
0 to 2 than are seen; (iii) the fact that the \Lya\ clouds are
particularly likely, in most simple theoretical pictures, to be
``ionized away'' when they are nearest to density peaks where quasars
(or other ionization sources) may have formed.  In fact, item (iii) may
be the explanation of item (ii); but it is clear that no simple
analysis can carry any degree of credibility.  We therefore defer these
issues to a later paper.

Here, sticking more closely to the data, it is of interest to see if
any evolution with redshift can be detected in the correlation
properties of the SSG sample.  Figure 7 also plots, as smaller open
symbols, the results of dividing the sample into 3 redshift ranges,
$z<3$, $3<z<3.5$, and $3.5<z$.  The corresponding dashed curves are
the predictions of the pure Poisson model, equation (\eqref{18}), with
no adjusted parameters and with the assumption of a Poisson distribution,
$\xi_S=0$.  In general, the
prediction for higher redshifts is towards larger fluctuations, due
both to the $\kappa$ dependence in equation (\eqref{30}) and to the
explicit dependence on $1+z$ in equation (\eqref{40}).  One sees that,
to the level of possible systematic errors previously described, there
is good agreement for the two higher redshift samples, with mean
redshifts $\left< z \right> = 3.3$ and $\left< z \right> = 3.8$.  For,
the lowest-redshift sample, with mean redshift $\left< z \right> =
2.8$, there is some indication (with large statistical error bars, as
shown in the Figure) of an excess variance -- the triangles (highest)
are about 2-$\sigma$ from the dotted (lowest) curve.  The excess has the
triangular instrumental profile, suggesting that it could be explained as
an unresolved correlation function (equation \eqref{39}) with
\be \int \xi(\lambda_{obs}) d\lambda_{obs} \approx (18\pm 9)
     \hbox{ \AA} \e{45} \ee
This value is several times larger than the value that we infer
from the data of Webb (1986) as shown in Carswell (1989); however the
error bars are large, so the discrepancy may not be real.

Qualitatively, our data support previous suggestions, e.g., by
Carswell on the basis of higher-redshift data from Atwood et al.
(1985) and Carswell et al.  (1987), that the correlation function
grows from an undetectable to a significant level in the fairly narrow
redshift range between $z\approx 3.5$ to $z\approx 2.5$.  (See also
Shaver, 1988.) This redshift range also marks the epoch where (i)
there is a change in the value of $\gamma$, the exponent relating
cloud density and redshift (see, e.g., Figure 9 in SSG), and (ii)
metal line clouds become important.  We think that it is fair game for
theorists to infer a connection among these phenomena.  In particular,
it may be that the spatially correlated population of \Lya\ clouds
that appears at redshifts $z\simless 2.8$, perhaps associated with
galaxy formation, is a completely distinct population from the more
rapidly disappearing (towards low redshifts) population of primordial,
spatially uncorrelated, clouds.

\section{Where Are All the Baryons?}\label{V}

Models for light element production in the hot big bang give good
limits on $\Omega_bH_0^2$, where $\Omega_b$ is the present fraction of
critical density due to baryonic matter.  Recent analyses (Olive et
al. 1990, Walker et al. 1991) give the value and uncertainty
\be \Omega_b h^2 = 0.013 \pm 0.003 \e{46}\ee

Obviously it is of interest to know whether this density of baryons
can be accommodated, at high redshift, completely in the observed
density of \Lya\ clouds.  By far the greatest uncertainty, as we shall
now see, lies in the ionization state of the clouds and, in turn,
in the question of whether the ionization state is consistent with
the interpretation of observed $b$ values as thermal, and/or with
an equilibrium thermal and ionization model.

The clouds are
predominantly ionized.  Measured, or deduced, column densities $N$
refer in the first instance to neutral hydrogen (whose corresponding
volume number density we will denote $n$), while the value of
$\Omega_b$ is determined by the corresponding column or volume
densities of total hydrogen, which we denote $N_H$ and $n_H$.
Let $f_0$ denote the hydrogen neutral fraction; and let $\rho$ denote
the total cosmological baryon density (including an assumed helium mass
fraction $\approx 0.25$); so we have the relations
\be N = f_0 N_H \qquad n = f_0 n_H \qquad \rho = 1.33 m_p n_H \e{47} \ee

We can compute the mean (smoothed) density of hydrogen $n_H$ at a
given redshift $z$ along the line of sight by knowing, as we do, the
mean density of clouds with each possible column density $N$ (equation
\eqref{18}), as follows:
\begin{eqnarray}  n_H &=& {dN_H\over dr_p} = {dz\over dr_p}{dN_H\over dz}
    \nonumber \\
    &=&  {dz\over dr_p} \int \left({N\over f_0(N)}\right) \NO (1+z)^\gamma
       N_{14}^{-\beta} dN_{14} \nonumber \\
    &=&  2.8\times 10^{-14} h \hbox{ cm}^{-3} (1+z)^{\gamma+2}
        (1+\Omega z)^{1/2} \int {N_{14}^{-\beta+1}\over f_0(N) } dN_{14} \e{51}
\end{eqnarray}
Here $r_p$ is physical (proper) distance (see equation \eqref{22a}), and
$f_0(N)$ denotes the harmonic mean value of $f_0$ for
all clouds with neutral column density between $N$ and $N+dN$,
averaging over any other internal parameters (notably $b$).
That is, $f_0(N)$ is the value defined by $N_H=N/f_0(N)$.
It is an important point that, if $f_0(N)$ is averaged correctly, then
equation (\eqref{51}) is {\em independent} of the physical size, shape,
and distribution of
the clouds, e.g., whether they are spherical, or thin sheets, clumped or
unclumped, etc.  Equation (\eqref{51}) requires only that a random
line of sight be a fair sample of the universe.

We can rewrite equation (\eqref{51}) as a direct measurement of
$\Omega_b h^2$,
\be \Omega_b h^2 = {1.33 m_p n_H\over (1+z)^3}\, {8\pi G\over 3
     \widehat{H}_0^2 } = 3.4\times 10^{-9} h (1+z)^{\gamma-1}
     (1+\Omega z)^{1/2} \int  {N_{14}^{-\beta+1}\over f_0(N) } dN_{14}
     \e{52} \ee
where $\widehat{H}_0 \defeq 100$ km s$^{-1}$.

\subsection{{\em Inconsistency of Equilibrium Model with $\Omega_b$}}

Let us consider first the baseline hypothesis that the clouds are in
thermal and ionization equilibrium with a background UV flux, and that
the measured $b$ velocity widths are thermal.  (We will see, in fact,
that this baseline hypothesis leads to an observational reductio ad
absurdum.)

It follows from the analysis of Black (1981) that there is a broad
regime of density $\rho$ and incident UV intensity $J_\nu$ (Lyman limit
intensity in units ergs cm$^{-2}$s$^{-1}$Hz$^{-1}$sr$^{-1}$) where both
the temperature of the gas, and also its neutral fraction, depend only
on the combination $J_\nu/\rho$.  In particular, if $\rho$ and $J_\nu$
are in c.g.s. units, then
\be T = 1690 \hbox{ K } \mu \left({J_\nu\over\rho}\right)^{2/7} \quad
  \hbox{or} \quad b = 5.3 \hbox{ km s}^{-1} \mu^{1/2}
   \left({J_\nu\over\rho}\right)^{1/7} \e{48} \ee
and
\be f_0 = 397 \left({J_\nu\over\rho}\right)^{-1.22} \e{49} \ee
(Here $\mu$ is the mean molecular weight of the ionized gas.)
According to Black, these expressions are applicable at least in the
range $5\times10^3 < T < 5\times 10^5$K, $10^{-6} < n_H < 10^{-3}
$cm$^{-3}$, $10^{-22} < J_\nu < 10^{-20}$ (c.g.s; note that Black's
$J_0$ is $4\pi$ times our $J_\nu$).

Equations (\eqref{48}) and (\eqref{49}) imply a unique relationship
between the neutral fraction $f_0$ and the velocity (or temperature)
parameter $b$,
\be f_0 = \left({b\over 8.56\hbox{ km/s}}\right)^{-8.54}
          \left({\mu\over 0.64}\right)^{4.27}
        = 3.7\times 10^{-6} \left({b\over 37\hbox{ km/s}}\right)^{-8.54}
          \left({\mu\over 0.64}\right)^{4.27}
\e{50} \ee
and a relationship among $n_H$, $b$, and $J_\nu$,
\be
n_H = {\rho\over 1.36 m_p} = 1.2\times 10^{-4}\hbox{ cm}^{-3}
     \left({b\over 37\hbox{ km/s}}\right)^{-7}
     \left({\mu\over 0.64}\right)^{7/2}
     \left({J_\nu\over 10^{-21} \hbox{ c.g.s}}\right)
\e{50a}\ee
(Hereafter,
we will take $\mu=0.64$ and suppress the parametric dependence on
$\mu$.)  Obviously the large values of the exponents in $b$ require
that we use these relations with some caution.  Various authors
have estimated $J_\nu$ as being in the range $10^{-22}$ to $10^{-21}$;
see Bajtlik, Duncan, and Ostriker (1988) for discussion.

A first application of equation (\eqref{52}) is to assume that $f_0$
is independent of $N$ and uniform throughout any given cloud.  This
would be true for the base case of pressure confined clouds in a
uniform confining medium and with a uniform UV illumination.  Then
there is no subtlety in averaging $f_0(N)$; it can come out of the
integral as $f_0$ in equation (\eqref{50}).  The remaining integral is
weakly divergent, so we must adopt some value $N_{max}$ for a cutoff
in the column density, in terms of which we get,
\be \Omega_b h^2 = 1.6 \times 10^{-3} h (1+z)^{\gamma-1}
     (1+\Omega z)^{1/2} \left( {b\over 37\hbox{ km s}^{-1}}\right)^{8.54}
     N_{14\,max}^{0.53} \e{53} \ee
Here the normalizing value for $b$, namely 37 km s$^{-1}$, has been
chosen to be the observed mean value for $b$ (Table 1).

One sees that the cosmologically observed value for $\Omega_b h^2$
(equation \eqref{46}) is not just easy to produce, it is easy to
vastly exceed: Table I shows that the appropriate average
$\left<b^{8.54}\right>^{1/8.54}$ is apparently on the order of 70
km s$^{-1}$; and $N_{14\,max}$ is surely not less than 100,
corresponding to $N_{max} = 10^{16}$cm$^{-2}$, and
probably greater.  These values would imply $\Omega_b h^2 \approx 5
h$, which is excluded even for a fully baryonic $\Omega=1$ universe!

One might momentarily wonder whether internal structure within a
single cloud, with the temperature $T$ varying along the line of
sight provides a possible resolution.  However, such variation always
acts in the wrong direction:  It is possible to hide {\em more} baryons
within a cooler, narrower, undetectable line core; but baryons implied by
an observed thermal line width must always be there.

One can turn the argument around by inverting equation (\eqref{53})
for $b$.  (Now we are on the good side of the large exponent, and thus
quite insensitive to the other assumptions made.)  One finds that, for
$N_{max}$ in the plausible range $10^{18}$cm$^{-2}$ down to
$10^{16}$cm$^{-2}$, the observed $\Omega_b$ is consistent with an
ionization fraction $f_0$ that {\em would in equilibrium derive} from
thermal values of $b$ in the range $20$ to $25$ km s$^{-1}$.  Looking
back at Figure 5, one sees that this velocity range is where the
observed $b$ distribution is falling off rapidly to smaller $b$
values; in other words, considering the possibility of observational
errors and other sources of dispersion, the deduced equilibrium $b$
resembles a lower bound to the observed $b$ values.

\subsection{{\em  Inconsistency of Equilibrium Model with Size and Mass
of Clouds}}

Let us follow the previous logic to the
extreme.  Equations (\eqref{50}) and (\eqref{50a}) give a characteristic
length scale $L$ for a cloud with parameters $N$, $b$, $J_\nu$, namely,
\be L \sim {N\over n_H f_0} \sim 7.3\times 10^{5}\hbox{ pc }
     \left({N\over 10^{15}\hbox{cm}^{-2}}\right)
     \left({b\over 37\hbox{ km/s}}\right)^{15.54}
     \left({J_\nu\over 10^{-21} \hbox{ c.g.s}}\right)^{-1}
\e{58}\ee
There are two separate points to be made about equation (\eqref{58}).
First, the value obtained for $L$ is implausibly large, not only by
the apparent factor of $\sim 100$ for the parameters having the
scaling values given (see discussion of cloud sizes at the end of
\S3.3), but also by an additional factor of about
$2^{15.54}\sim 4\times 10^4$ if the distribution for $b$ has the broad
tail found in \S3.

\def\beq{b_{\rm eq}}
In view of the large exponent on $b$, a better use of
equation (\eqref{58}) is, as before, to run it backwards:
Suppose that $L$ is in
the range $10^{4\pm 1}$pc, as seems required by other observations, and
that the other parameters have the scaling values given.  Then (\eqref{58})
implies
\be \left<\beq^{15.54}\right>^{1/15.54} \approx 28 \pm 5\hbox{ km s}^{-1}
 \e{59} \ee
where $\beq$ is the thermal $b$ value that would be in equilibrium
with the required neutral fraction $f_0$.

The point can be made even more
forcefully if we write the characteristic mass $M$ of the cloud (more
accurately, the mass of that part of the cloud whose size is on the
order of the line of sight penetration),
\be M \sim \rho L^3 \sim 1.48 \times 10^{12} M_\odot
     \left({N\over 10^{15}\hbox{cm}^{-2}}\right)^3
     \left({b\over 37\hbox{ km/s}}\right)^{39.62}
     \left({J_\nu\over 10^{-21} \hbox{ c.g.s}}\right)^{-2}
\e{60}\ee
Here the value of the exponent in $b$ is truly worthy of awe.  If we
reverse the equation, and assume $M$ in the range $10^{8\pm 3} M_\odot$,
which is cosmologically plausible, we get
\be \left<\beq^{39.62}\right>^{1/39.62} \approx 29 \pm 6\hbox{ km s}^{-1}
 \e{61} \ee

An additional interesting point about equation (\eqref{60}) is that,
because of the enormous exponent, it is simply not possible to
accomodate {\em any} significant range of $b$ within the
cosmologically available mass range for $M$, if $b$ is related to
neutral fraction by equation (\eqref{53}) or anything similar.  A
factor of two variation in $b$ from cloud to cloud, at fixed $N$,
induces a mass range of a factor $10^{12}$!
Also note that the observed range of $N$,
which we take conservatively in this paper to be $\sim 10^3$, itself
induces a mass range of a factor $10^{9}$.  Moreover, many observers (e.g.,
Petitjean, 1992) show evidence of $N$ itself extending over 8 orders
of magnitude!

\section{Possible Resolutions}\label{VI}

Quite obviously, the UV-driven equilibrium thermal model for $b$ is
unviable.  On the one hand (\S 3), a broad distribution of $b$'s,
extending to 80 km s$^{-1}$ or higher,  is observationally required.
On the other hand (\S 6), the neutral fractions implied by such
values of $b$ are incompatible with observed limits on $\Omega_b$,
the size and mass of the clouds, and the dynamic range of mass
available to the clouds.

Any possible resolution of the paradox must substantially disconnect a
measured value of $b$ from the ionization state of its cloud.
There would seem to be two ways to do this.  First, as has been
long noted in the literature, the observed value $b$ could be due
to bulk matter motions rather than thermal velocities.

The bulk motion hypothesis raises a range of theoretical difficulties
that have been discussed by other investigators.  More
observationally, in the context of this paper, it is worth noting
that, if $b$ includes a significant component of bulk motions, then
any curve of growth analysis based on Voigt profiles (including that
of \S2) will be very seriously in error in the saturated region.  In
particular, if, within a single cloud, the velocity tail falls off
less rapidly than a Gaussian (as one would expect of the broad-tailed
distributions characteristic of hydrodynamic motions), then the actual
value of $N$, the neutral column density, can be much less than
deduced from standard curve of growth.  The widely observed number
distribution of $N$'s up to high values is then based on an incorrect
analysis, and the fact that the observed distribution is a power law
that extrapolates smoothly from smaller $N$ values becomes something
of a miracle.

One should also mention the
observational issue usually termed ``the $b-N$ controversy'',
and described in Shaver, Wampler, and Wolfe (1991).  There, claims
is of a positive observed correlation between $b$ and $N$ (which would
be theoretically expected in most mechanisms for bulk motion) have
proved quite difficult to substantiate and are thought by many to
be entirely due to selection effects.  In this context we should
note the success of our assumption, in \S 3, of uncorrelated $N$'s
and $b$'s in satisfying all observational constraints considered.

Finally, it is hard to reconcile most mechanisms for generating
bulk motion with the observed lack of cloud clumping on small scales
(\S 4).  Hydrodynamic or gravitational processes capable of driving
highly supersonic differential motions within a cloud should also be
expected to act on scales comparable to intercloud distances
(\S 3.3).  Our analysis of fluctuations in \S 4 would easily have
detected (e.g. in the zero-lag bin of Figure 7) any tendency for clouds
to be clumped in groups as small as a few, or for that matter any
comparable tendency for clouds to be more uniformly distributed than
random.  The lack of such a detection tends to support less violent
scenarios of cloud formation and evolution than those that can give
large bulk motions.

An alternative to the bulk motion hypothesis (as particularly
emphasized to us by M. Rees; see also Rees 1988) is the notion that
observed $b$'s do represent the thermal cloud state, but that this
thermal state is not in equilibrium with the ionization state
determined by interaction with the UV background.  In particular, it
may be possible to place the clouds in a physical regime where their
thermal cooling times are not short compared with the (then) Hubble
time.

If such a picture is combined with the idea that we are seeing clouds
that have collapsed and heated adiabatically, then a consistent
picture may possibly emerge:  The neutral fraction $f_0$ is
determined over time by interaction with a background UV flux.
There is an implied concomitant universal heating of the gas, to
a temperature that one might identify with a value at or below
the lower edge of the $b$ distribution in Figure 5.

Subsequently, clouds collapse by volume factors that vary somewhat
from cloud to cloud, giving the observed broad, indeed thermal,
distribution of $b$.  Since adiabatic heating is inefficient at
altering ionization, the values $f_0$ remain in a universal, narrow
range, as required by the arguments of \S 5.  In Paper III we will
investigate this, and related, models.

\section{Conclusions}\label{VII}

The conclusions of this paper are as follows:

1.  The equivalent width ratios for the Lyman sequence found in Paper
I are completely consistent with a featureless power law distribution
of \Lya\ clouds in column density $N$, with exponent $\beta = 1.43\pm
0.04$.  Distributions with a break in the power law can also fit the
equivalent width ratio data when the break is around
$N=10^{15}$cm$^{-2}$.

2.  Assuming the above featureless power law distribution in $N$,
standard curve of growth analysis is able to reproduce the detailed
distribution of cloud equivalent widths (notably both the exponential
tail at large equivalent widths, and the turn-up at small equivalent
widths), but only for distributions in $b$ that have mean values $\sim
37$ km s$^{-1}$ and a significant tail extending as high as 70 km s$^{-1}$.
Such distributions are in fact observed by some observers.

3.  The turn-up at small equivalent widths likely does not mark any
change in the properties of the clouds, but is a consequence purely
of the curve of growth along the observed line of sight.

4.  Broken power laws, which do fit the $N$ distribution, do not fit
the equivalent width distribution as well, but they are not wholly ruled
out.

5.  The long-observed exponential tail in the distribution of equivalent
widths can be real without being physically fundamental: it emerges as
an artifact of combining the tail of the distribution in $b$ with the
slowly rising equivalent widths characteristic of the curve of growth
in the saturated region.

6.  The posited distributions in $N$ and $b$,
plus Paper I's normalization of the absorption as a function of redshift,
give an absolute normalization on the number of clouds as a function of
$N$, $b$, and $z$.  Upper limits on the physical size of clouds,
in the range of 50 to 120 $h^{-1}$ kpc at $z\sim 4.2$, follow.

7.  At the highest redshifts studied, the posited distributions in $N$
and $b$ are able completely to account for fluctuations in the
absorption (along different lines of sight or as a function of
redshift), as due to Poisson fluctuations in an uncorrelated cloud
distribution.  However, there are marginal (not by themselves very
believable) detections of a nonzero correlation function in two
different regimes: (i) At the lowest redshift accessible to this
study, $z\approx 2.8$, there is some evidence of unresolved
correlation on a scale $\simless 2 h^{-1}$ Mpc, possibly with
amplitude somewhat greater than previously reported.  (ii) In the full
sample, on scales of tens of Mpc, there is some evidence of
correlation at a level $\xi \sim 0.01$ or $0.02$.  Because of possible
systematic errors, however, we prefer, to take twice these values as
upper limits.

8. If the observed $b$ distribution is thermal, and if the ionization
state of the gas is related to its thermal state by an equilbrium
model (e.g., Black 1981), then the observed broad $b$ distribution
would imply $\Omega_b \gg 1$ in clouds.  Similarly, the broad $b$
distribution implies clouds that are both too large and too massive.

9.  On the other hand, if the ionization state is that implied by a
UV-heated temperature of about 20 to 25 km s$^{-1}$, then at high
redshift the \Lya\ clouds contain all the baryons deduced from models
of light element production in the hot big bang (i.e., probably all
the baryons in the universe).

10.  The broad $b$ distribution must therefore be ascribed either
to bulk motions (which raises both theoretical difficulties and
likely incompatibilities both with the observed featureless power law
distribution in $N$ obtained from curve-of-growth analyses,
and with the observed lack of small-scale cloud clumping), or else
the observed clouds must be out of thermal equilibrium, and
heated by a process that is relatively inefficient at ionization.
Recent adiabatic collapse is a plausible candidate.

\acknowledgments

We have benefited from discussions with Martin Rees, Don Schneider,
John Bahcall, David Spergel, John Black, Ed Turner, and John Huchra.
This work was supported in part by the National Science Foundation
(PHY-91-06678).

\clearpage

\begin{figure}
\caption{Equivalent width $W$ for Lyman series lines in hydrogen as
a function of column density $N$ and thermal velocity parameter $b$.
Equivalent widths grow linearly with $N$ in the unsaturated region
below $N\sim 10^{14-15}$ cm$^{-2}$, until they saturate at a value
set by the thermal doppler width.  In the saturated regime, widths
grow slowly as $\sim\log^{1/2} N$.}
\end{figure}

\begin{figure}
\caption{Goodness-of-fit $\chi^2$ values as a function of exponent $\beta$
for fitting observed Lyman series absorption ratios and a curve of
growth model to a power law model for the number of clouds as a
function of their column density $N$.  With four measured ratios,
there are four degrees of freedom.  The value of the exponent is seen
to be narrowly determined (note expanded abscissa) independently of
the assumed value for the velocity parameter $b$.}
\end{figure}

\begin{figure}
\caption{Goodness-of-fit $\chi^2$ values as in Figure 2, but fitting
for the characteristic column density $N$ associated with
the broken power-law model of equation (13) [broad parabolas]
or the unrealistic model of a single, universal $N$ [narrow parabolas].
One sees that, while none of these models are excluded by consideration of
absorption ratios alone, the ratios are fairly powerful at fixing
one scale parameter.}
\end{figure}

\begin{figure}
\caption{Relative number of \Lya\ clouds as a function of their rest
equivalent width. Filled circles: data compiled by Murdoch et
al.~(1986).  Note the famous exponential tail extending over $>2$
decades in cloud number density.  Solid, long- and short-dashed
curves: Predictions based on curve of growth analysis and the three
distributions for velocity parameter $b$ that are shown in Figure 5.
The exponential tail emerges naturally from $b$ distributions with
$\left< b \right> \approx 37$ km s$^{-1}$ and $\sigma_b
\approx 20$ km s$^{-1}$ (see text
for details).  Dotted and dash-dot curves: Models which deviate from
{\it either} a broad $b$ distribution {\it or} a pure power-law
distribution in $N$ do not give acceptable fits to the observed
equivalent width distribution.}
\end{figure}

\begin{figure}
\caption{Solid histogram: Carswell's (1989) compilation of the distribution
of $b$ values for \Lya\ clouds.  Long- and short-dashed curves: gamma
law and truncated Gaussian distributions, each parametrized by a mean and
width, which give best fits to the observed equivalent width distribution
shown in Figure 4.  One sees that the behavior of the $b$ distribution
at small values of $b$ is not well constrained, but that both models,
and Carswell's data, indicate a tail of large $b$ values extending to
$> 80$ km s$^{-1}$.}
\end{figure}

\begin{figure}
\caption{To determine the range of column densities $N$ that are
tested by the excellent agreement of models with data in Figure 4, the
mean value $N$ of clouds with a specified equivalent width is here
plotted against that equivalent width, for four of the $b$ models
shown in Figure 4.  One sees that all the models probe the range
$N\sim 10^{13}$ cm$^{-2}$ to $N\sim 10^{16}$ cm$^{-2}$.}
\end{figure}

\begin{figure}
\caption{Fluctuations in transmission along the line of sight, here
shown as the fractional covariance of absorption along a single line
of sight measured at two wavelengths separated by an integral number
of 10\AA\ bins.  Solid dots: observed covariance for the full SSG
sample of quasars. Error bars are determined by bootstrap resampling.
Solid curve: prediction (with no adjustable parameters) for
unclustered clouds with a distribution in $z$, $N$, and $b$ determined
in this paper without reference to fluctuation statistics.  The
triangular shape comes from the instrumental response of an assumed
square slit.  The good agreement between the data and the prediction
put strong limits on a cloud correlation function $\xi$.  The shoulder
in bins 3 through 7, if real, implies a value $\xi \sim 0.02$ on a
comoving scale $\sim 10\, h^{-1}$Mpc at redshift $z\approx 3.5$.
However, systematic errors in the determination of the zero point are
also of the same order, so the detection is suggestive only. (See text
for details.)  The open symbols and dotted curves repeat the analysis
for low-, medium-, and high-reshift subsamples of the full sample.
There is weak (about 2 $\sigma$) evidence for the appearance of a
positive correlation function in the lowest-redshift sample (see
text).}
\end{figure}

\end{document}